\documentclass[a4paper,12pt]{article}
\usepackage{graphicx}
\newcommand{\beq}{\begin{equation}}
\newcommand{\eeq}{\end{equation}}
\newcommand{\bea}{\begin{eqnarray}}
\newcommand{\eea}{\end{eqnarray}}
\newcommand{\bsub}{\begin{subequations}}
\newcommand{\esub}{\end{subequations} \noindent}

\def\PRD#1#2#3{Phys. Rev. {\bf D#1} (19#2) #3}

\def\NPB#1#2#3{Nucl. Phys. {\bf B#1} (19#2) #3}

\def\ZPC#1#2#3{Z. Phys. {\bf C#1} (19#2) #3}
\def\EPJC#1#2#3{Eur. Phys. J. {\bf C#1} (19#2) #3}
\def\PLB#1#2#3{Phys. Lett. {\bf B#1} (19#2) #3}
\def\PRL#1#2#3{Phys. Rev. Lett. {\bf #1} (19#2) #3}

\makeatletter
\newtoks\@stequation

\def\subequations{\refstepcounter{equation}%
  \edef\@savedequation{\the\c@equation}%
  \@stequation=\expandafter{\theequation}
  \edef\@savedtheequation{\the\@stequation}
  \edef\oldtheequation{\theequation}%
  \setcounter{equation}{0}%
  \def\theequation{\oldtheequation\alph{equation}}}

\def\endsubequations{%
  \ifnum\c@equation < 2 \@warning{Only \the\c@equation\space subequation
    used in equation \@savedequation}\fi
  \setcounter{equation}{\@savedequation}%
  \@stequation=\expandafter{\@savedtheequation}%
  \edef\theequation{\the\@stequation}%
  \global\@ignoretrue}

\def\eqnarray{\stepcounter{equation}\let\@currentlabel\theequation
\global\@eqnswtrue\m@th
\global\@eqcnt\z@\tabskip\@centering\let\\\@eqncr
$$\halign to\displaywidth\bgroup\@eqnsel\hskip\@centering
     $\displaystyle\tabskip\z@{##}$&\global\@eqcnt\@ne
      \hfil$\;{##}\;$\hfil
     &\global\@eqcnt\tw@ $\displaystyle\tabskip\z@{##}$\hfil
   \tabskip\@centering&\llap{##}\tabskip\z@\cr}

\makeatother

\begin{document}
\thispagestyle{empty}
\vspace*{-15mm}
\baselineskip 10pt
\begin{flushright}
\begin{tabular}{l}
{\bf KEK-TH-671}\\
{\bf SNS-PH/00-02}\\
{\bf hep-ph/0001229}
\end{tabular}
\end{flushright}
\baselineskip 24pt 
\vglue 10mm 
\begin{center}
{\Large\bf
Muon $g-2$ and precision electroweak physics \\
in the MSSM }
\vspace{8mm}

\baselineskip 18pt 
\def\thefootnote{\fnsymbol{footnote}}
\setcounter{footnote}{0}
{\bf Gi-Chol Cho$^{1),2)}$, Kaoru Hagiwara$^{1)}$ and 
Masashi Hayakawa$^{1)}$}
\vspace{5mm}

$^{1)}${\it Theory Division, KEK, Tsukuba, Ibaraki 305-0801, Japan}\\
$^{2)}${\it Scuola Normale Superiore, Piazza dei Cavalieri 7, 
I-56126 Pisa, Italy}\\
\vspace{10mm}
\end{center}
\begin{center}
{\bf Abstract}\\[7mm]
\begin{minipage}{12cm}
\baselineskip 16pt
\noindent
The minimal supersymmetric extension of standard model (MSSM)
is examined by analyzing its quantum effects on the precision 
electroweak measurements and the muon $g-2$.
We examine carefully the effects of light charginos and neutralinos 
that are found to improve the fit to the electroweak data. 
We identify two distinct regions on the $(\mu, M_2)$-plane that 
fit well to the electroweak data and give significant contribution 
to muon $g-2$.
\end{minipage}
\end{center}
\newpage
\baselineskip 18pt 
 The minimal supersymmetric standard model (MSSM) has been one of 
the extensively investigated theories beyond the standard model 
since it was recognized~\cite{Hagiwara} that it has the potential 
to unify strong and electroweak gauge interactions.
 On the other hand, despite enormous efforts to find its signature, 
there has been no direct evidence for the existence of the 
supersymmetric particles. 
 In this letter we study quantitatively the MSSM predictions for 
the muon $g-2$ in the light of the latest precision electroweak 
data~\cite{lepewwg98,wboson_moriond}, and that of the present 
and future $g-2$ experiments~\cite{future-exp}.

The importance of the muon $g-2$, conventionally denoted as 
$a_\mu = \frac{1}{2} (g_\mu - 2)$, has been widely discussed in 
the context of supersymmetric 
theories~\cite{Kosower,Chattopadhyay,Polonsky,Gabrielli}. 
The observation of the effect from the MSSM is found generally 
accessible with the target accuracy of the current experiment at 
Brookhaven National Laboratory (BNL)~\cite{future-exp}
\begin{equation}
  \Delta a_\mu({\rm expt}) =
  4.0 \times 10^{-10}\, .
   \label{eq:future-precision}
\end{equation}
This accuracy is about $1/20$ of the error (numerals in the 
parenthesis) of the current measurement~\cite{future-exp}
\begin{equation}
a_\mu(\textrm{expt}) = 11659~235~(73) \times 10^{-10}\, .
\label{eq:present_value}
\end{equation}
The prediction of the standard model (SM) is
\begin{equation}
a_\mu({\rm SM}) = 11659\ 168.75\ (9.56) \times 10^{-10}\, ,
\label{eq:sm_value}
\end{equation}
where the details of specific contributions is found in 
Ref.~\cite{HK}. 
We shall come back to the theoretical uncertainties of the SM 
prediction not included in the above quoted error of 
$9.56 \times 10^{-10}$ in the summary part.

 As noted above the experimental results on the precision 
measurements around the $Z$-pole have played an essential role 
in revealing the possibility of grand unification.
This implies the existence of many supersymmetric particles below 
the TeV scale and hence we may expect to observe their quantum 
effects in the precision electroweak experiments as well as in 
the muon $g-2$.
 Although no signal of supersymmetry has so far been identified, 
recent systematic study of the precision electroweak data found 
that the existence of the relatively light ($\sim$ 100 GeV) 
charginos and neutralinos can make the MSSM fit slightly better 
than that of the standard model~\cite{CH}. 
 We therefore study carefully the MSSM effects on the muon $g-2$ 
in the range $\left| \mu \right|$, $M_2$ $<$ 500 GeV. 
 It is found that analysis performed here could give rise to 
a systematic criterion to identify the preferred range of the 
MSSM parameters.


\vspace{0.5cm}
 As we will see, the additional contribution to the muon $g-2$ in 
the MSSM can be greater than the target accuracy 
(\ref{eq:future-precision}) of the present experiment. 
 Thus it is convenient to define the additional new physics 
contribution to the muon $g-2$ in the unit of $10^{-10}$
\begin{equation}
 \delta a_\mu \equiv
  10^{10} \times
   \left\{
    a_\mu({\rm MSSM}) - a_\mu({\rm SM})
   \right\} \, .
  \label{eq:a_scaled}
\end{equation}
 The present experimental data (\ref{eq:present_value})
can then be expressed as
\begin{equation}
 -16 \le
 \delta a_\mu \le 149\, ,
  \label{eq:g-2_bound}
\end{equation}
at the 1-$\sigma$ level.
 Here we add the theoretical uncertainty of 10 
in the unit of $10^{-10}$ in eq.~(\ref{eq:sm_value}) 
linearly to the present experimental error of 73
in eq.~(\ref{eq:present_value}).
%

In the MSSM $a_\mu $ receives essentially two new contributions. 
One comes from the chargino ($\widetilde{\chi}^-_j, j=1,2$)
and muon-sneutrino ($\widetilde{\nu}_\mu$) propagation in the 
intermediate states, and the other from the neutralino 
($\widetilde{\chi}^0_j, j=1\sim 4$) and smuon 
($\widetilde{\mu}_i, i=1,2$) intermediate states.
The chargino-sneutrino contribution can be expressed as
\bsub
\begin{eqnarray}
a_\mu(\widetilde{\chi}^-) &=& 
\frac{1}{8\pi^2}\,\frac{m_\mu}{m_{\widetilde{\nu}_\mu}}
\sum_{j=1}^2
\left\{
\frac{m_\mu}{m_{\widetilde{\nu}_\mu}}\,
G_1 \left(
     \frac{m_{\widetilde{\chi}^-_j}^2}{m_{\widetilde{\nu}_\mu}^2}
            \right)
\left(
     \left| g_L^{\widetilde{\chi}^-_j \mu \widetilde{\nu}_\mu} 
          \right|^2
     +
\left| g_R^{\widetilde{\chi}^-_j \mu \widetilde{\nu}_\mu} \right|^2
\right)
 \right.
\\
 && \quad \quad \quad \quad \quad \quad
 +
 \left.
   \frac{m_{\widetilde{\chi}^-_j}}{m_{\widetilde{\nu}_\mu}}\,
   G_3 \left(
        \frac{m_{\widetilde{\chi}^-_j}^2}{m_{\widetilde{\nu}_\mu}^2}
       \right)
   {\rm Re}
     \left[
      \left(
       g_R^{\widetilde{\chi}^-_j \mu \widetilde{\nu}_\mu}
      \right)^*
      g_L^{\widetilde{\chi}^-_j \mu \widetilde{\nu}_\mu}
     \right]
 \right\} , \nonumber \\
 G_1(x) &=& 
  \frac{1}{12(x-1)^4}
  \left[
   (x-1)(x^2 - 5x - 2) + 6 x\,{\rm ln}\,x
  \right] \, , \\
 G_3(x) &=& 
  \frac{1}{2(x-1)^3}
  \left[
   (x-1)(x-3) + 2\,{\rm ln}\,x
  \right]\, ,
\end{eqnarray}
  \label{eq:chg-snr-mdm}
\esub
while the neutralino-smuon contribution as
\bsub
\begin{eqnarray}
 a_\mu(\widetilde{\chi}^0) &=&
 -
 \frac{1}{8\pi^2}
 \sum_{i=1}^2 \frac{m_\mu}{m_{\widetilde{\mu}_i}}
 \sum_{j=1}^4
 \left\{
   \frac{m_\mu}{m_{\widetilde{\mu}_i}}
   G_2\left(
       \frac{m_{\widetilde{\chi}_j^0}^2}{m_{\widetilde{\mu}_i}^2}
      \right)
   \left(
     \left| g_L^{\widetilde{\chi}^0_j \mu \widetilde{\mu}_i} 
          \right|^2
     +
     \left| g_R^{\widetilde{\chi}^0_j \mu \widetilde{\mu}_i} 
          \right|^2
   \right)
 \right.
 \nonumber \\
 && \quad \quad \quad \quad \quad \quad \quad \quad
 \left.
  + \frac{m_{\widetilde{\chi}_j^0}}{m_{\widetilde{\mu}_i}}\,
    G_4\left(
        \frac{m_{\widetilde{\chi}_j^0}^2}{m_{\widetilde{\mu}_i}^2}
       \right)
    {\rm Re}
    \left[
      \left(
       g_R^{\widetilde{\chi}^0_j \mu \widetilde{\mu}_i}
      \right)^*
      g_L^{\widetilde{\chi}^0_j \mu \widetilde{\mu}_i}
    \right]
 \right\} , \\
 G_2(x) &=&
  \frac{1}{12(x-2)^4}
  \left[
   (x-1)(2x^2 + 5x - 1) - 6 x^2\,{\rm ln}\,x
  \right]\, , \\
 G_4(x) &=&
  \frac{1}{2(x-1)^3}
  \left[
   (x-1)(x+1) - 2 x\,{\rm ln}\,x
  \right]\, .
\end{eqnarray}
 \label{eq:ntr-smu-mdm}
\esub
 Here we adopt the notation of Ref.~\cite{CH,susy_lagrangian} 
for the coupling constants 
\begin{equation}
 {\cal L} = \sum_{\alpha = L, R}
            g_\alpha^{F_1 F_2 S} \overline{F}_1 P_\alpha F_2 S\, ,
\end{equation}
where $F_1$ and $F_2$ are four-component fermion fields, 
$S$ denotes a scalar field, and
\begin{equation}
 P_L = \frac{1-\gamma_5}{2}\, ,\quad
 P_R = \frac{1+\gamma_5}{2}\, .
\end{equation}
 The charged Higgs boson contribution, which gives rise to a sizable 
effect in the case of $b \rightarrow s \gamma$ transition, is highly 
suppressed due to the small Yukawa couplings in the muon $g-2$. 
 The MSSM contribution to the muon $g-2$ is most significant at large 
$\tan\beta$~\cite{Kosower}.
 The chargino-sneutrino loop, eq.~(\ref{eq:chg-snr-mdm}), then 
dominates over the other in almost all region of the $(\mu, M_2)$ 
plane.
 However the neutralino-smuon loop contribution to $\delta a_\mu$ 
is also larger than the target accuracy (\ref{eq:future-precision}) 
of the current experiment. 
 Except for $(\mu, M_2, \tan\beta)$ which determines the chargino 
and neutralino masses, the former depends on the left-handed SUSY 
breaking slepton mass, $m_{\widetilde{L}}$, while the latter also 
depends on the right-handed one, $m_{\widetilde{E}}$, in addition. 
 We discuss this point in more detail when we study the dependence
of $\delta a_\mu$ on $m_{\widetilde{E}}$.

In our analysis we allow the MSSM parameters to vary freely restricted 
only by experimental constraints from direct and indirect searches.
 When restricted to specific models of supersymmetry (SUSY) breaking 
such as the minimal supergravity scenario, or the gauge mediated 
supersymmetry breaking scenario, we reproduce the known 
results~\cite{Chattopadhyay,Gabrielli}. 
 First we consider the case when all the squarks are so heavy that 
their effects on low-energy physics disappears.
 In our actual numerical calculation, 
we set all the squark masses and 
the extra Higgs boson masses at 1 TeV 
such that they do not make worse the fit to the electroweak
data \cite{CH,Precision_theory}
and the $b\to s\gamma$ rate \cite{b-s-gamma_th}. 

 The constraints on the MSSM parameter space from the electroweak 
experiments have been found in Ref.~\cite{CH} by using the 18 $Z$-pole 
data given by the LEP/SLC experiments~\cite{lepewwg98} and the 
$W$-boson mass $m_W$ given by the Tevatron and LEP2 
experiments~\cite{wboson_moriond}. 
 The electroweak observables are affected by the supersymmetric 
particles through both the universal gauge boson propagator 
corrections and the process specific vertex/box corrections. 
 Following the formalism introduced in Ref.~\cite{CH}, 
the universal part of the radiative corrections of the $Z$-pole 
observables can be represented by two oblique parameters $S_Z$ and 
$T_Z$, while the $W$-boson mass $m_W$ itself can be adopted as the third 
oblique parameter. 
 The parameters $S_Z$ and $T_Z$ are expressed as the sum of 
the conventional $S$ and $T$ parameters~\cite{peskin-takeuchi,HHKM} 
and the $R$ parameter which measures the running effect of the 
$Z$-boson propagator correction between $q^2=0$ and $q^2=m_Z^2$. 
 The current electroweak data favors new physics that gives negative 
contribution {\it both} to the $S_Z$ {\it and} $T_Z$ parameters, 
and hence they constrain strongly the additional positive 
contributions to these parameters in the MSSM.
The squark contributions to the $T_Z$ parameter are generally found to 
be positive while they do not affect to the $S_Z$ parameter so much. 
 The light charginos and neutralinos are found to give negative 
contributions to the $R$ parameter and hence they make both 
the $S_Z$ and $T_Z$ parameters negative, thereby improving the fit 
to the electroweak data. 
 Light left-handed sleptons make $S_Z$ negative but keep $T_Z$ 
essentially unchanged. 
 It has been shown that if the chargino mass is close to its lower 
mass bound from the LEP2 experiment, 
$m_{\widetilde{\chi}^-_1} \sim 100$ GeV, 
and the squarks and sleptons are sufficiently heavy, say 1 TeV, 
the total $\chi^2$ in the MSSM is slightly better than that in the 
standard model~\cite{CH}: 
\begin{equation}
 \Delta \chi^2 \equiv 
  \chi^2_{\rm MSSM} - \chi^2_{\rm SM} \sim -1\, .
  \label{eq:Delta_chi}
\end{equation}
No other combinations of light supersymmetric particles are found to 
improve the fit to the electroweak data over the standard model. 

 Summing up, the precision electroweak data favors light charginos 
and neutralinos but disfavors light left-handed squarks and sleptons. 
 The muon $g-2$ is found to be sensitive to the left-handed slepton 
mass when there are light charginos and neutralinos. 
 If the left-handed sleptons are relatively light, then the MSSM 
contribution to $g-2$ grows but the electroweak fit worsens. 
 Below we examine quantitatively the MSSM prediction for the muon 
$g-2$ and the electroweak data, and look for the region of the MSSM 
parameter space which gives observable $g-2$ effect without spoiling 
the good fit to the electroweak data. 

\begin{figure}[t]
\begin{center}
 \includegraphics[width=7cm,clip]{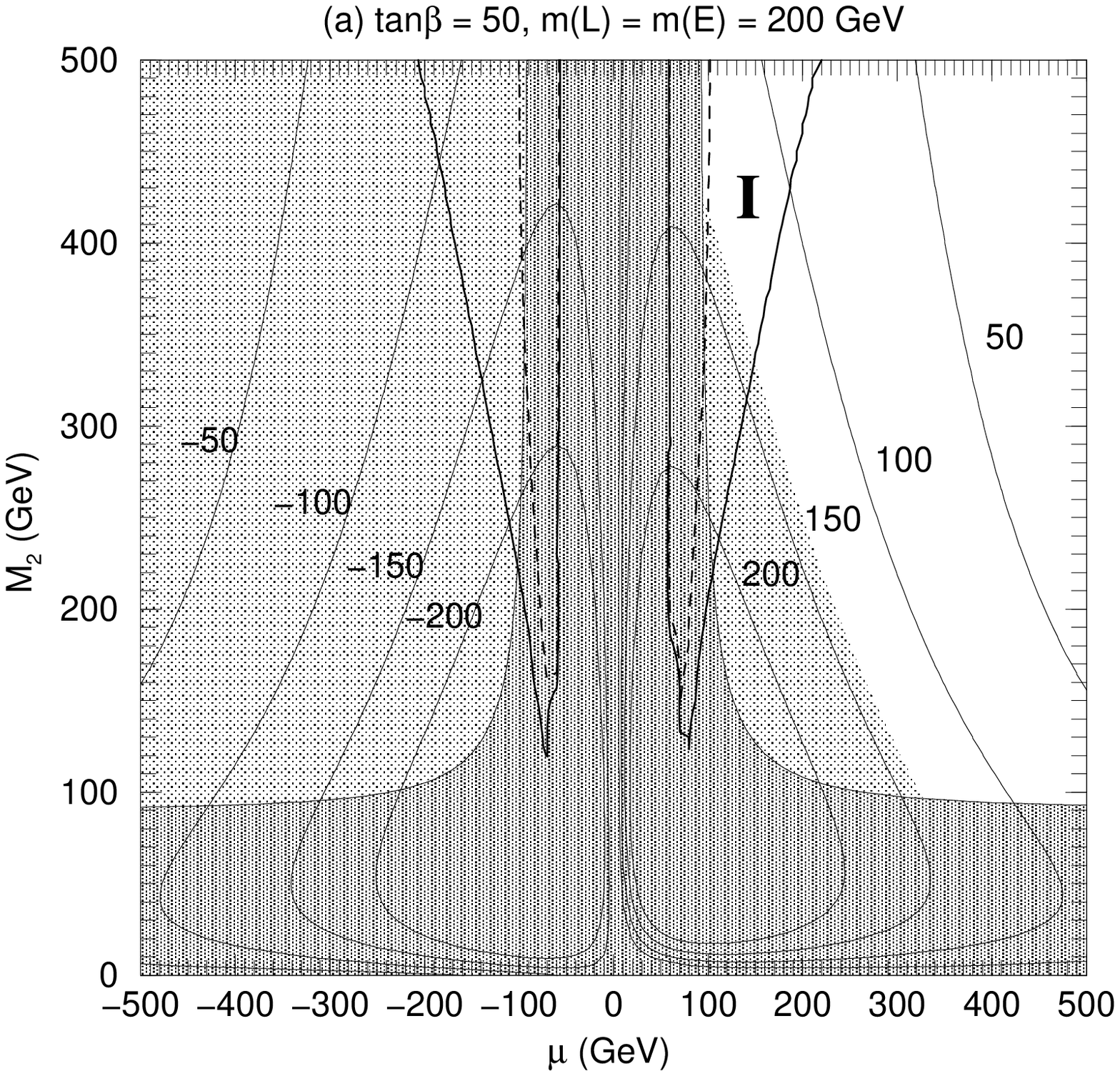}
 \includegraphics[width=7cm,clip]{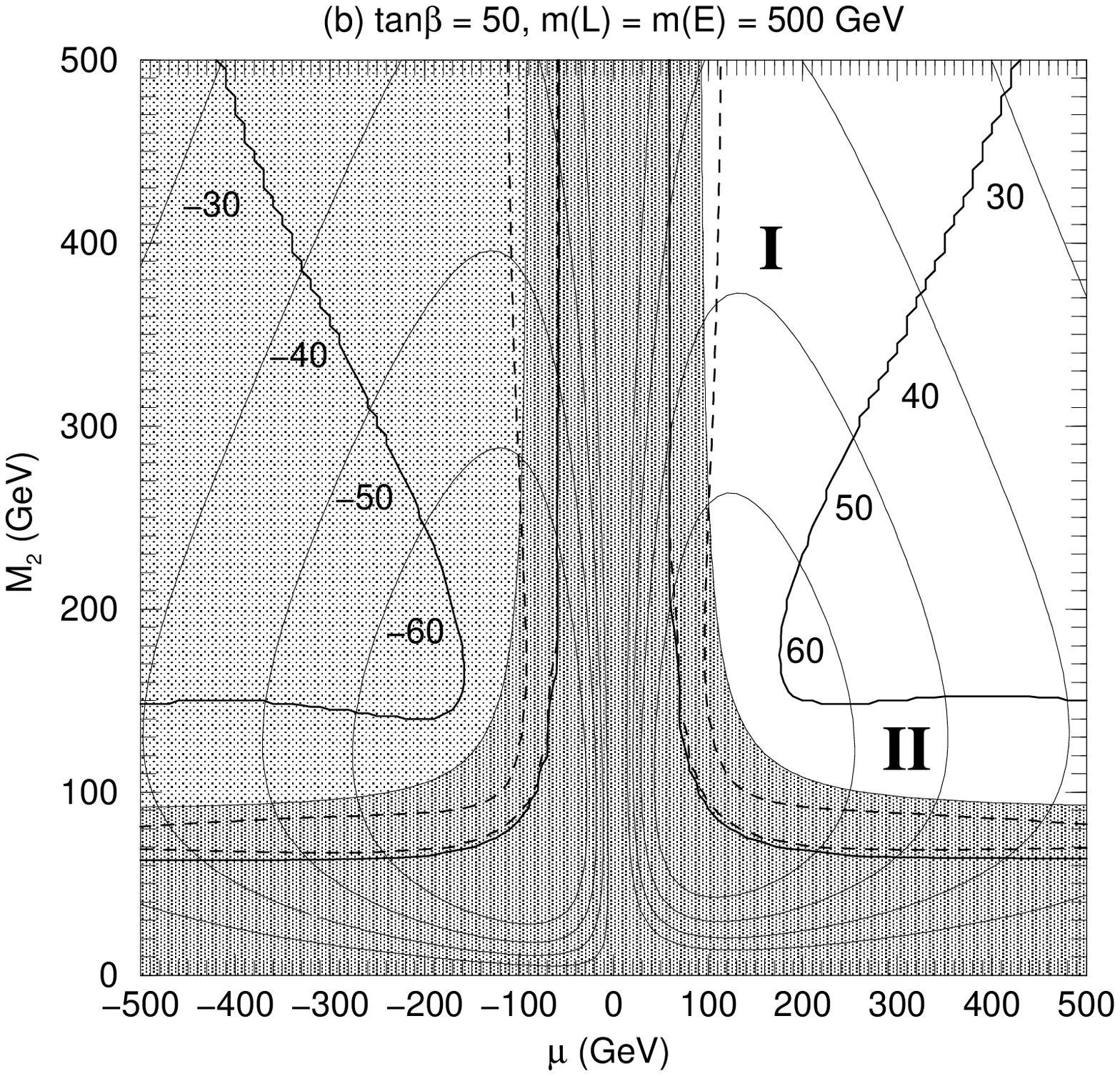}
 \caption{
  Contours of the MSSM contribution
 to the muon $g-2$ in the unit of $10^{-10}$ ($\delta a_\mu$)
 on the $(\mu, M_2)$-plane for $\tan\beta=50$.
 (a) $m_{\widetilde{L}}$ = $m_{\widetilde{E}}$ = 200 GeV.
 (b) $m_{\widetilde{L}}$ = $m_{\widetilde{E}}$ = 500 GeV.
  The region enclosed by the dashed line
 gives $\Delta \chi^2 = \chi^2_{\rm MSSM} - \chi^2_{\rm SM}$ 
 $<$ 0 for the precision electroweak data,
 while the region enclosed by the bold line
 satisfies $\Delta \chi^2$ $<$ 1.
  The darkly shaded zone is excluded
 from the current lower bound on the chargino mass.
  The lightly shaded region 
 is excluded by the currently allowed value for muon $g-2$
 in eq. (\ref{eq:g-2_bound}).
 I and II on the graphs indicate
 Region I in Eq. (\ref{eq:region_I}) 
 and Region II in Eq. (\ref{eq:region_II}) respectively.}
  \label{fig:tan50}
\end{center}
\end{figure}
%
 Taking this circumstances in mind, we show in Fig. \ref{fig:tan50} 
the contours of fixed values of $\delta a_\mu$ defined in 
Eq. (\ref{eq:a_scaled}) on the ($\mu$, $M_2$)-plane for a relatively 
large value of $\tan\beta = 50$. 
 We set $m_{\widetilde{L}}$ = $m_{\widetilde{E}}$ = 200 GeV in 
Fig.~\ref{fig:tan50}(a) and $m_{\widetilde{L}}$ = $m_{\widetilde{E}}$ 
= 500 GeV in Fig.~\ref{fig:tan50}(b), respectively. 
 For the gaugino masses, we adopt the ``unification'' condition, 
$M_1/M_2$ = $\alpha_1(M_Z) /\alpha_2(M_Z)$, for simplicity. 
 However, the general aspect of our conclusion obtained here persists 
as long as the gaugino masses share a common order of magnitude. 
The shaded-region shows the region of the parameters already 
inconsistent with the mass bound for chargino~\cite{Mihara}
\begin{equation}
 m_{\widetilde{\chi}^-_1} > 93\ {\rm GeV}\, .
\end{equation}
 The physical slepton masses are obtained as the eigenvalues 
of the slepton mass-squared matrix for each flavor 
($l$ = $e$, $\mu$ or $\tau$) : 
\bsub
\bea
 &&
 M_{\widetilde{l}}^2 =
  \left(
   \begin{array}{cc}
    M_{LL}^2 & M_{LR}^2 \\
    M_{RL}^2 & M_{RR}^2
   \end{array}
  \right)\, ,
  \\
 &&
 M_{LL}^2 = m_{\widetilde{L}}^2 + m_l^2
           - M_Z^2 \cos 2\beta
             \left(
              \frac{1}{2} - \sin^2 \theta_W
             \right)\, ,
  \\
 &&
 M_{RR}^2 = m_{\widetilde{E}}^2 + m_l^2
           - M_Z^2 \cos 2\beta \sin^2 \theta_W\, ,
  \\
 &&
 M_{LR}^2 = (M_{RL}^2)^* =
   m_l (A^*_l - \mu \tan\beta)\, ,
\eea
  \label{eq:slp_mass}
\esub
where $m_l$ denotes the charged-lepton mass. 
 In Fig. \ref{fig:tan50} we show the contours of muon $g-2$
when $A_\mu$ is set equal to zero.
 Effects of nonzero $A_\mu$ are discussed below.

\begin{figure}[t]
\begin{center}
 \includegraphics[width=7cm,clip]{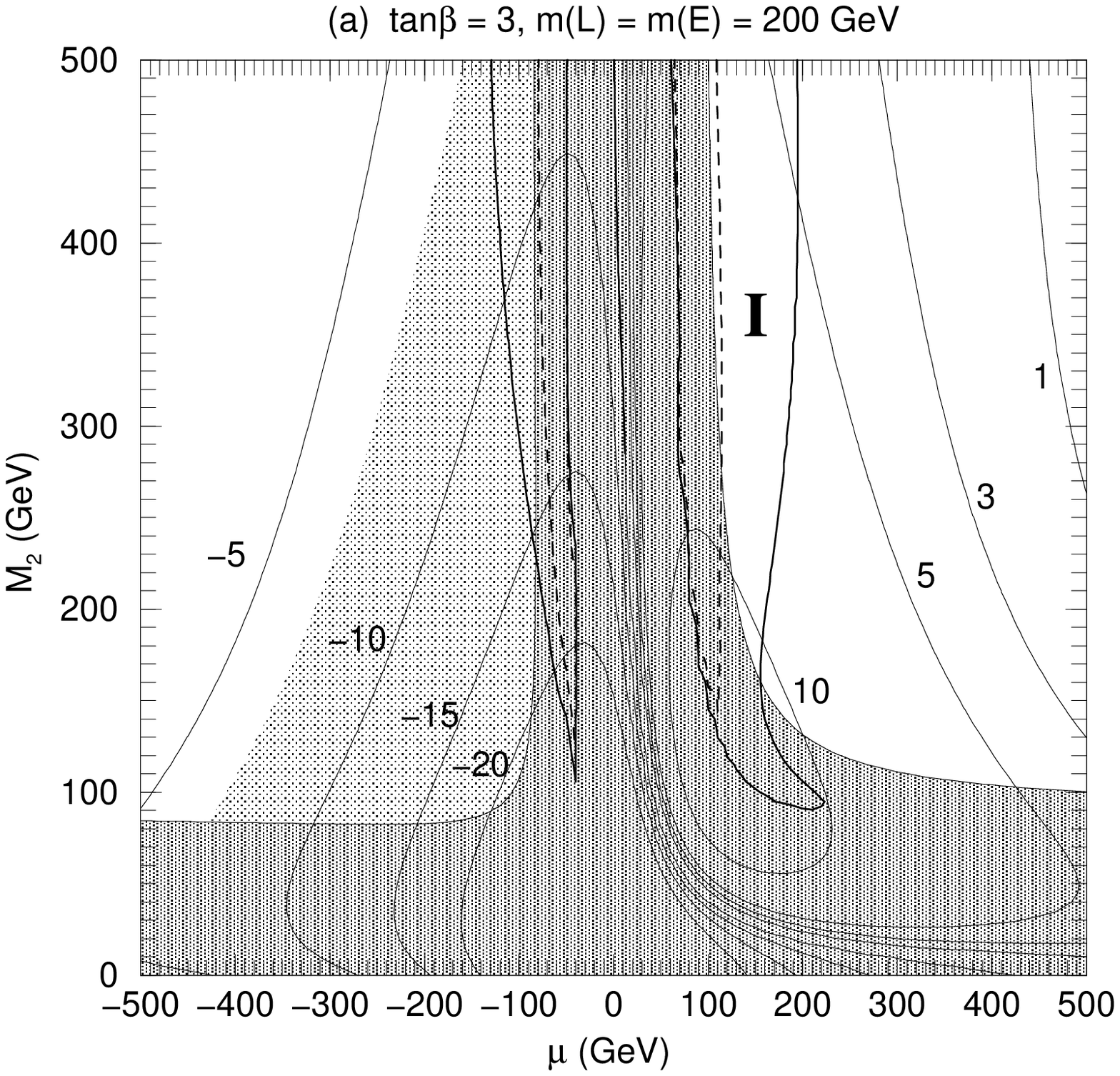}
 \includegraphics[width=7cm,clip]{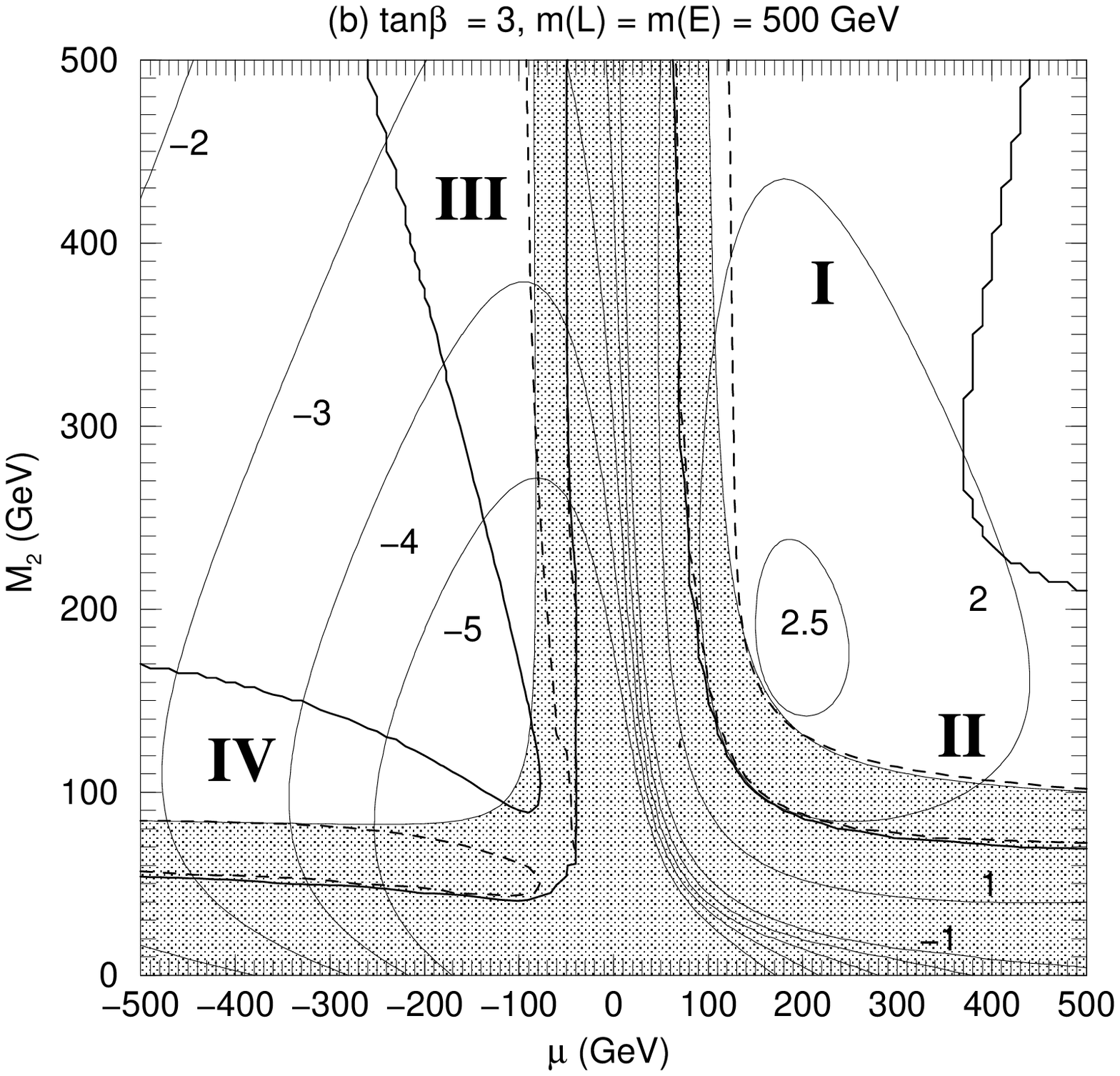}
 \caption{
  Graph similar to {\protect Fig. \ref{fig:tan50}}
  for $\tan\beta=3$.
  (a) $m_{\widetilde{L}}$ = $m_{\widetilde{E}}$ = 200 GeV.
  (b) $m_{\widetilde{L}}$ = $m_{\widetilde{E}}$ = 500 GeV.
  III and IV on the graphs indicate
  Region III in Eq. (\ref{eq:region_III})
  and Region IV in Eq. (\ref{eq:region_IV}) respectively.
 }
  \label{fig:tan3}
\end{center}
\end{figure}

 Fig.~\ref{fig:tan50}(a) shows a generic feature that 
the MSSM contributions can be as large as 
$(100 \sim 300) \times 10^{-10}$ when $\tan\beta$ = 50
and $m_{\widetilde{L}}$ = 200 GeV, which is much greater than 
the $W$- and $Z$-contributions in the SM~\cite{Czarnecki},
\begin{equation}
 a_\mu(\textrm{weak}) = 15.1~(4) \times 10^{-10}\, .
  \label{eq:weak_effect}
\end{equation}
 This is essentially because of the large Yukawa coupling for 
the $\mu_R \widetilde{h}^- \widetilde{\nu}_\mu$ vertex at large 
$\tan\beta$ $\gg$ 1.
 Here $\mu_R$ is the right-handed muon component and 
$\widetilde{h}^-$ is the charged component of the higgsino 
which couples to charged leptons and down-type quarks. 
 The magnitude of the MSSM predictions reduces to the level of 
the SM weak corrections (\ref{eq:weak_effect}) 
at around $\tan\beta = 3$, as shown in Fig. \ref{fig:tan3}. 
 The MSSM contribution to the muon $g-2$ scales roughly as 
$r_{\widetilde{L}} /m_{\widetilde{L}}^2 
 + r_{\widetilde{E}} /m_{\widetilde{E}}^2$ 
in the large $\tan\beta$ case, where $r_{\widetilde{L}}$ and 
$r_{\widetilde{E}}$ are some constants. 
 The dependence on $m_{\widetilde{L}}$ and $m_{\widetilde{E}}$ 
will be further evaluated below.

 Also shown in the figures is the region where the fit to 
the electroweak data is better than that of the SM 
(the narrow region enclosed by the dashed lines). 
 The region falls mostly into the region forbidden by 
the chargino mass bound (the darkly shaded regions). 
 The region enclosed by the bold lines 
gives reasonably good fit to the electroweak data, 
where $\Delta \chi^2$ in Eq. (\ref{eq:Delta_chi}) is less than one. 
 We find that there are essentially two regions 
which respect the result 
from the precision electroweak measurements. 
 One region is
\begin{equation}
 {\rm Region\,\,I}\,:\,
  \quad 120\ {\rm GeV} < \mu < c_1 \, M_2,
  \label{eq:region_I}
\end{equation}
where $c_1$ is a positive number greater than one, 
depending on the slepton masses. 
 The other minor region appears only 
when $m_{\widetilde{L}}$ is taken to be 
larger than about 500 GeV (see Fig. \ref{fig:tan50}(b)):
\begin{equation}
 {\rm Region\,\,II}\,:\,
  100\ {\rm GeV} < M_2 < 150\ {\rm GeV}\, .
  \label{eq:region_II}
\end{equation}
 The lightest chargino is generally higgsino-dominated 
in Region I 
while it is gaugino-dominated in Region II.
 Both domains correspond to the extreme regime 
in which the chargino mass becomes nearly equal to 
the experimental lower bound, around 
which the chargino-neutralino contributions 
to the oblique parameters pull both $S_Z$ and $T_Z$ back to 
negative directions. 
 This negative contribution to $\Delta \chi^2$ 
is necessary to complement the positive contribution 
(which worsens the fit) due to relatively small $m_{\widetilde{L}}$. 
 Thus the MSSM with large $\tan\beta$ 
gives a sizable contribution to the muon $g-2$ 
if $\left| \mu \right|$, $M_2$ and $m_{\widetilde{L}}$ 
are all smaller than 500 GeV. 
 Actually the present muon $g-2$ data (\ref{eq:g-2_bound})
already excludes the negative $\mu$ region 
of Fig.~\ref{fig:tan50}.
%
%

\begin{figure}[t]
\begin{center}
 \includegraphics[width=7cm,clip]{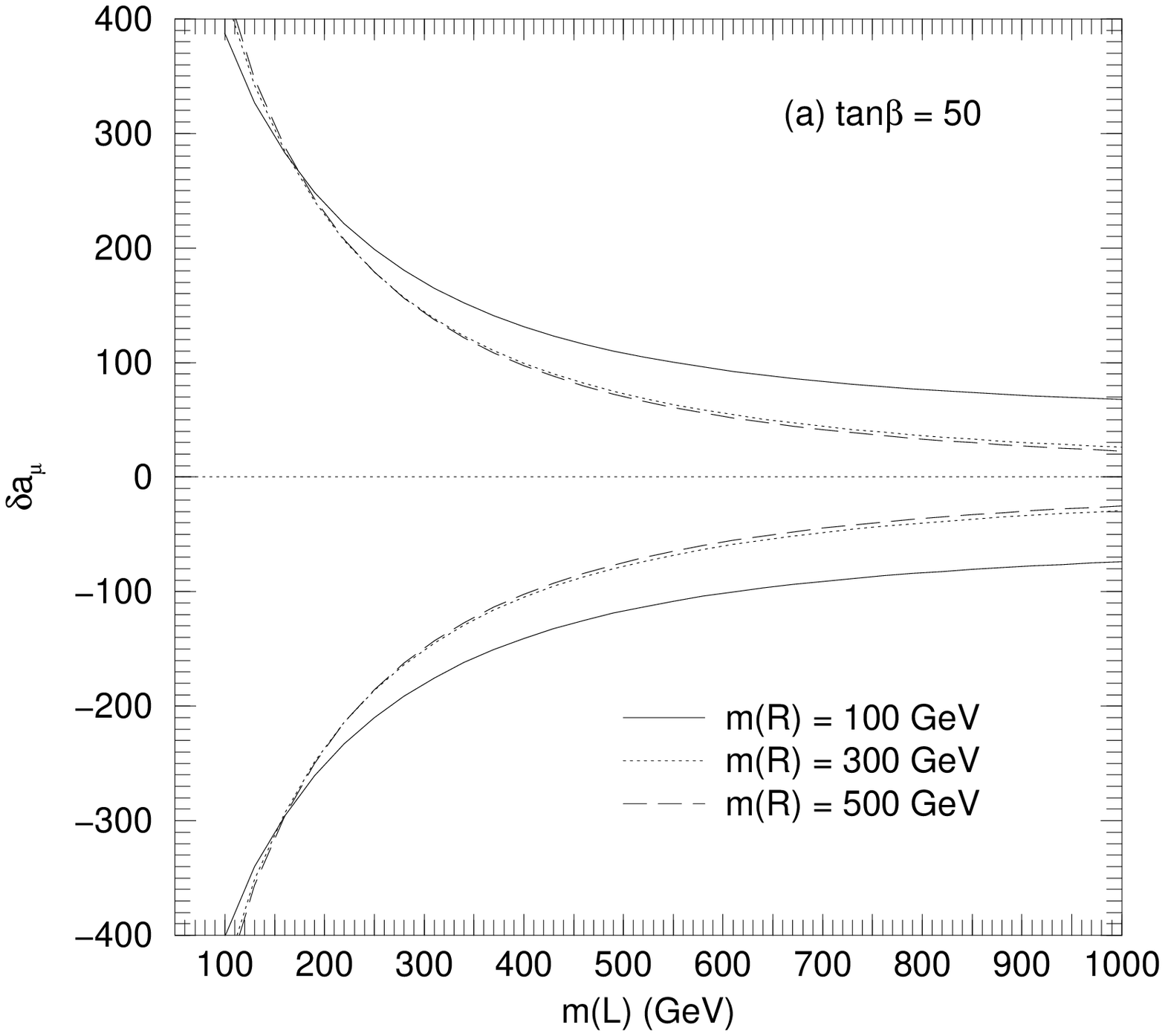}
 \includegraphics[width=7cm,clip]{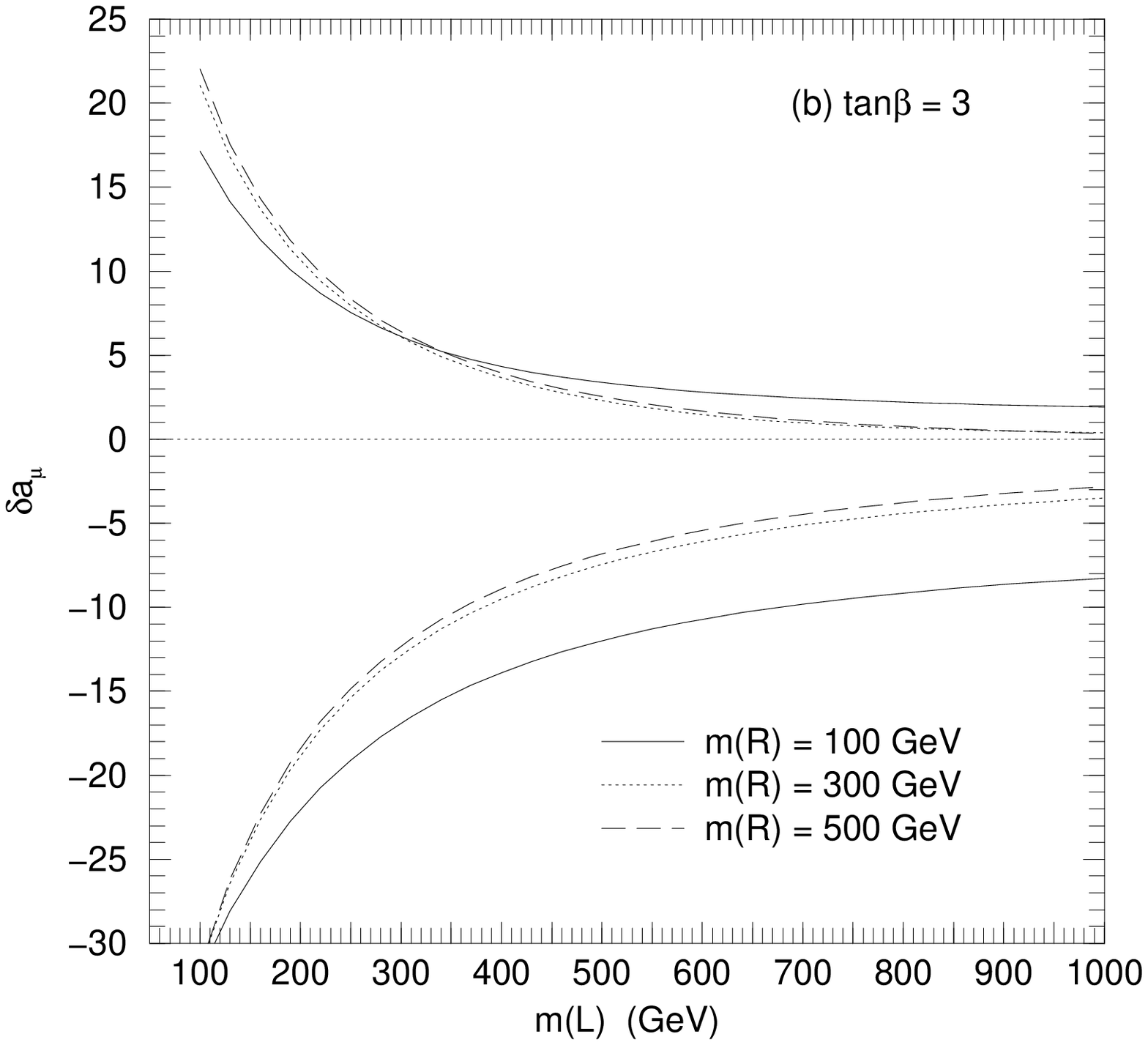}
 \caption{
  $m_{\widetilde{L}}$-dependence of the maximally 
  possible MSSM contributions to the muon $g-2$ 
  in the unit of $10^{-10}$ ($\delta a_\mu$)
  for (a) $\tan\beta=50$ and (b) $\tan\beta = 3$
  when $A_\mu = 0$. 
  The right-handed slepton mass $m_{\widetilde{E}}$ 
  is taken to be 100 GeV (solid line), 300 GeV (dotted line) 
  and 500 GeV (dashed line). 
 }
  \label{fig:slp_dep}
\end{center}
\end{figure}

 The muon $g-2$ is a less powerful probe of the MSSM 
at smaller $\tan\beta$. 
 Fig.~\ref{fig:tan3} shows the similar graphs
as Fig.~\ref{fig:tan50}, but for $\tan\beta = 3$. 
 As in Fig.~\ref{fig:tan50}(a) and (b),
$m_{\widetilde{L}}$ = $m_{\widetilde{E}}$ = 200 GeV 
in Fig.~\ref{fig:tan3}(a), and
$m_{\widetilde{L}}$ = $m_{\widetilde{E}}$ = 500 GeV
in Fig.~\ref{fig:tan3}(b).
 Here we are interested in 
whether we can find some evidence of the MSSM 
with small $\tan\beta$ 
through the measurement of the muon $g-2$ 
in view of the other precise low energy data.
 Fig.~\ref{fig:tan3}(a) shows that the region with positive $\mu$ 
and small $M_2$ for 
small slepton mass ($m_{\widetilde{L}}$ = 200 GeV) 
is accessible by the present muon $g-2$ experiment with 
its target precision~(\ref{eq:future-precision}), 
while respecting the result of the electroweak experiments. 
 Fig.~\ref{fig:tan3}(b) shows that
new two allowed regions appear on the negative $\mu$ side
of the $(\mu, M_2)$-plane
for slepton masses larger than about 500 GeV.
 Those two regions are
\bsub
\begin{eqnarray}
 &&
 {\rm Region\ III}:\quad -c_2\, M_2 < \mu < -80\ {\rm GeV}\, , 
  \label{eq:region_III} \\
 &&
 {\rm Region\ IV}: \quad 80\ {\rm GeV} < M_2 < -c_3\, \mu \ ,
                  \quad \mu < -120\ {\rm GeV}\ ,
  \label{eq:region_IV} 
\end{eqnarray}
\esub
where $c_2$ and $c_3$ are positive numbers 
depending on $m_{\widetilde{L}}$ and $m_{\widetilde{E}}$. 
 The MSSM correction to the muon $g-2$ is small compared to 
the target precision~(\ref{eq:future-precision}) 
in those regions in Fig.~\ref{fig:tan3}(b). 
 However, when $m_{\widetilde{E}}$ $<$ $m_{\widetilde{L}}$,
the observable enhancement might be expected 
in Region III and IV,
as will be shown in Fig.~\ref{fig:slp_dep}(b). 

 Next we study the MSSM prediction to the muon $g-2$ 
without assuming the universality between 
$m_{\widetilde{L}}$ and $m_{\widetilde{E}}$. 
 In Fig.~\ref{fig:slp_dep} we show 
the $m_{\widetilde{L}}$-dependence of 
the maximally admissible $\delta a_\mu$ 
for (a) $\tan\beta = 50$ and (b) $\tan\beta = 3$
when $A_\mu = 0$. 
In each figure, the solid, dotted and dashed lines represent 
$m_{\widetilde{E}}=100$ GeV, 300 GeV and 500 GeV, respectively. 
 We find that, for $\tan\beta$ = 50, the maximum 
of $\delta a_\mu$ is achieved when $M_2 /\mu$ is nearly 
equal to $\pm 1$. 
 For $\tan\beta$ = 3, this property also holds as long as 
$m_{\widetilde{L}}$ is small enough (less than about 500 GeV).

 It should be remarked that 
the precision measurements around $Z$-pole 
favor large left-handed slepton mass 
but is rather blind to the right-handed slepton mass. 
 Fig.~\ref{fig:slp_dep} tells us that 
the MSSM prediction of the muon $g-2$
has sizable $m_{\widetilde{E}}$-dependence
when the target accuracy (\ref{eq:future-precision})
of the current $g-2$ experiment
is taken into account.
 Let us recall that 
the chargino-sneutrino loop contribution 
(\ref{eq:chg-snr-mdm}) depends on
$m_{\widetilde{L}}$
while the neutralino-smuon loop contribution 
(\ref{eq:ntr-smu-mdm}) depends on both
$m_{\widetilde{L}}$ and $m_{\widetilde{E}}$.
 The numerical study shows that 
the chargino-sneutrino loop contribution 
decreases faster than the neutralino-smuon loop contribution 
if $m_{\widetilde{L}}$ is increased 
while $m_{\widetilde{E}}$ is kept fixed. 
 Thus $\delta a_\mu$ is more sensitive to relatively small 
$m_{\widetilde{E}}$ for larger $m_{\widetilde{L}}$. 
 This is because only the neutralino-smuon loop contribution 
depends on $m_{\widetilde{E}}$ and 
the lighter smuon is almost 
the right-handed slepton component 
for $m_{\widetilde{E}}$ $\ll$ $m_{\widetilde{L}}$. 
 The neutralino-smuon loop gives nonzero contribution to 
$\delta a_\mu$ even for $m_{\widetilde{L}} \rightarrow \infty$. 
 This explains why $\delta a_\mu$ does not decouple
in Fig.~\ref{fig:slp_dep} at large $m_{\widetilde{L}}$.
 Indeed, for $\tan\beta$ = 50, $m_{\widetilde{E}}$ = 100 GeV 
and $m_{\widetilde{L}}$ $\rightarrow$ $\infty$, 
the maximally possible $\delta a_\mu$ is 65, 
which is observable in the light of the precision 
(\ref{eq:future-precision}) for the muon $g-2$ measurement. 
 Fig.~\ref{fig:slp_dep} shows that, 
on account of the constraint on $\delta a_\mu$ 
(\ref{eq:g-2_bound}), 
there is a possibility that 
the MSSM with small $\tan\beta$ 
for negative $\mu$ and $m_{\widetilde{E}}$ $<$ $m_{\widetilde{L}}$ 
can be probed by the current muon $g-2$ experiment with 
the target precision (\ref{eq:future-precision}). 

 So far, all our results have been shown 
for $A_\mu$ = 0. 
 We discuss the dependence of these results 
on $A_\mu$ here. 
 We first recall that 
$A_\mu$ enters only in the left-right mixing 
of the smuon mass-squared matrix~(\ref{eq:slp_mass}) 
for all the electroweak precision observables 
and the muon $g-2$ at one-loop order. 
 Thus $A_\mu$-dependence comes from 
the neutralino-smuon loop contribution. 
 The numerical analysis 
shows that $\delta a_\mu$ varies at most 5 
when $A_\mu$ varies from $-$1 to 1 TeV
for the various choices of $m_{\widetilde{L}}$, 
$m_{\widetilde{E}}$, $M_2$ and $\mu$
in the allowed region at both the large and small $\tan\beta$. 
Thus our findings in Fig.~\ref{fig:tan50} and 
Fig.~\ref{fig:slp_dep}(a) remain valid up to the target 
precision (\ref{eq:present_value}) 
of the current muon $g-2$ experiment
even with nonzero $A_\mu$. 
 In particular the slepton mass dependence 
does not change quantitatively. 
 For small $\tan\beta$, the impact of nonzero $A_\mu$ is 
comparable to the magnitude of $\delta a_\mu$ in $A_\mu=0$ case. 
 Then, our discussion on the small $\tan\beta$ case with 
$A_\mu=0$ might be modified depending on the magnitude and 
the sign of $A_\mu$. 
 Note that, if $\tan\beta$ is small, $A_\mu$ could affect 
$\delta a_\mu$ when the relative sign between $A_\mu$ and $\mu$ 
is opposite. 
 This is because $A_\mu$ always appears as a linear 
combination with $\mu\tan\beta$ 
in the slepton mass-squared matrix element (\ref{eq:slp_mass}). 
 We find that $\delta a_\mu$ for positive $\mu$ could be at most 
15 if $A_\mu$ is negative and large, say $\sim -1$ TeV. 

\vspace{0.5cm}
 To summarize the paper we performed the quantitative analysis 
on the MSSM contribution to the muon $g-2$ 
on account of the constraint 
from the precision electroweak data. 
 The analysis \cite{CH} on the precision 
electroweak measurements 
tells us that the fit of the MSSM might be better than 
that of the SM when the chargino mass is 
close to the LEP2 bound ($\sim$ 100 GeV) 
and the left-handed sfermions are heavy enough, 
say a few hundred GeV. 
 Our study has been done by taking account of this point
and the target precision $4 \times 10^{-10}$ in 
(\ref{eq:present_value}) of the current muon $g-2$ experiment. 

 At first, assuming the universality between the left- and the 
right-handed SUSY breaking mass parameters, 
$m_{\widetilde{L}} = m_{\widetilde{E}}$, 
we found four regions on the $(\mu, M_2)$-plane 
where the fit to the electroweak data is not worse 
as compared to the SM 
and the predicted MSSM contribution 
to the muon $g-2$ is allowed from the current experimental result. 
 We found that the muon $g-2$ is significantly enhanced 
(of the order 100 $\times 10^{-10}$) in Region I 
(120 GeV $<$ $\mu$ $<$ $c_1 M_2$, 
with $c_1$ $>$ 1) for $\tan\beta$ = 50 
and $m_{\widetilde{L}}$ = $m_{\widetilde{E}}$ = 
200 GeV (see Fig.~\ref{fig:tan50}(a)). 
 For the heavy slepton mass, $m_{\widetilde{L}}$ = 
$m_{\widetilde{E}}$ = 500 GeV, 
and $\tan\beta$ = 50,
Region II (100 GeV $<$ $M_2$ $<$ 150 GeV) appears in addition, 
where the predicted value of the muon $g-2$ 
is allowed in the current experiment 
but the MSSM correction is 
at most about 60 $\times$ $10^{-10}$
in both Region I and II. (See Fig.~\ref{fig:tan50}(b).) 

 We also examined the dependence 
on $m_{\widetilde{L}}$ and $m_{\widetilde{E}}$ 
of those results without universality assumption. 
 We found that
$m_{\widetilde{E}}$-dependence 
of the muon $g-2$ is larger than 
its target precision of the present experiment 
in the large $\tan\beta$ case. 
 Thus the muon $g-2$ complements the precision electroweak 
measurement which is rather blind to $m_{\widetilde{E}}$. 
 It might be a unique possible indirect probe 
to give constraints on the right-handed slepton.
 When $m_{\widetilde{E}}$ $<$ $m_{\widetilde{L}}$, 
the MSSM with small $\tan\beta$ is 
also accessible by the current $g-2$ experiment
(See Fig.~\ref{fig:slp_dep}(b)). 
 The model falls in Region III (\ref{eq:region_III}) 
or in Region IV (\ref{eq:region_IV}) 
on the $(\mu, M_2)$-plane for $\mu < 0$. 
 It is also found that 
the MSSM contribution to the muon $g-2$ 
is always affected by at most $5 \times 10^{-10}$ 
from the $A_\mu$-term. 
 There is no essential change in the above results
in the large $\tan\beta$ case. 
 For negative $A_\mu$ of the order 1 TeV, 
the MSSM with the small $\tan\beta$ 
in Region I and II, rather than Region III and IV,  
can be probed by the muon $g-2$ measurement.
 The detectable limit of the chargino at Tevatron Run-II is 
expected to be 250 GeV \cite{Tevatron}. 
 If the chargino is found at Tevatron, 
the muon $g-2$ will enable us to select 
the parameters of the MSSM .

 Here we remind the reader that 
the allowed range (\ref{eq:g-2_bound}) for $\delta a_\mu$ 
is derived from (\ref{eq:present_value}) 
and (\ref{eq:sm_value}). 
 If the result in Ref.~\cite{Davier}, 
which utilizes the accurate data 
for hadronic decay of $\tau$ 
(See Ref.~\cite{Kuhn} and the references therein)
further with the help of quark-hadron 
duality in evaluation of dispersion integral, is applied instead 
of the result in Ref.~\cite{Alemany}, 
the leading order QCD correction 
involved in the above quantity receives $(8\sim9) \times 10^{-10}$ 
modification, resulting in 
\begin{equation}
 a_\mu({\rm SM}) =
  11659\ 16 0.05\ (6.44) \times 10^{-10}\, .
  \label{eq:sm_value_2}
\end{equation}
 This could then change the allowed range
of $\delta a_\mu$ from (\ref{eq:g-2_bound}) to
\begin{equation}
 -4 < \delta a_\mu < 154\, .
  \label{eq:g-2_bound_2}
\end{equation}
In so far as we illustrate the tendency of the excluded region 
at large $\tan\beta$ with the current experimental 
value~(\ref{eq:present_value}),
the difference between 
(\ref{eq:g-2_bound}) and (\ref{eq:g-2_bound_2}) does not matter.
But such a difference will become crucial to find appropriate sets 
of ($\mu$, $M_2$) when the experimental uncertainty is reduced 
to the level (\ref{eq:future-precision}). 
This aspect is quite interesting from the view point of 
searching the MSSM at large $\tan\beta$. 
Thus the reduction of the error residing in QCD corrections to 
the muon $g-2$ is a necessary task to be put forward promptly. 
\vspace{0.5cm}
\\
{\bf \Large Acknowledgment}
\vspace{0.3cm}
\\
 The authors thank S. Eidelman for 
 the various usuful comments on the manuscript. 
\newpage

%

\begin{thebibliography}{99}
%
 \bibitem{Hagiwara}
  K.~Hagiwara,
   Ann.~Rev.~Nucl.~Part.~Sci. {\bf 48} (1998) 463.
%
 \bibitem{lepewwg98}
  The LEP Collaborations ALEPH, DELPHI, L3, OPAL, 
  the LEP Electroweak Working Group and the SLD Heavy 
  Flavor Group, CERN-EP/99-15. 
%
 \bibitem{wboson_moriond} 
  I.~Riu, talk given at the XXXIVth Rencontres de Moriond, 
  March 13-20, 1999. 
 \bibitem{future-exp}
  R.~M.~Carey {\it et. al},
  \PRL{82}{99}{1632}. 
%
 \bibitem{Kosower}
  J.~Ellis, J.~Hagelin and D.~V.~Nanopoulos,
  \PLB{116}{82}{283}; \\
  J.~A.~Grifols and A.~Mendez,
  \PRD{26}{82}{1809}; \\
  D.~A.~Kosower, L.~M.~Krauss and N.~Sakai,
  \PLB{133}{83}{305}. 
%
 \bibitem{Chattopadhyay}
  U.~Chattopadhyay and P.~Nath,
    \PRD{53}{96}{1648}.
%
 \bibitem{Polonsky}
  J.~L.~Lopez, D.~V.~Nanopoulos and X.~Wang,
    \PRD{49}{94}{366}; \\
  T.~Moroi,
    \PRD{53}{96}{6565};
    {\it ibid}. {\bf D56}, 4424 (1997) (E); \\
  F.~Borzumati, G.~R.~Farrar, N.~Polonsky and S.~Thomas,
 	\NPB{555}{99}{53}.    
%
 \bibitem{Gabrielli}
  M.~Carena, G.~F.~Giudice and C.~E.~Wagner,
   \PLB{390}{97}{234}; \\
  E.~Gabrielli and U.~Sarid,
   \PRL{79}{97}{4752};
   \PRD{58}{98}{115003};
%
 \bibitem{HK}
  M.~Hayakawa and T.~Kinoshita,
   \PRD{57}{98}{465};
%
 \bibitem{CH}
  G.~C.~Cho and K.~Hagiwara,
   hep-ph/9912260 (1999), to appear in Nucl.~Phys.~{\bf B}.
%
 \bibitem{susy_lagrangian}
  G.~C.~Cho and K.~Hagiwara,
    ``The MSSM Larangian for MadGraph2'', in preparation. 
%
%
 \bibitem{Precision_theory}
  D.~M.~Pierce, J.~A.~Bagger, K.~Matchev and R.~Zhang,
     \NPB{491}{97}{3}; \\
  W.~de Boer, A.~Dabelstein, W.~Hollik, W.~Mosle and 
    U.~Schwickerath,
    \ZPC{75}{97}{627}; \\
  P.~H.~Chankowski, J.~Ellis and S.~Pokorski,
     \PLB{423}{98}{327}; \\
  J.~Erler and D.~M.~Pierce,
    \NPB{526}{98}{53}; \\
  I.~V.~Gaidaenko, A.~V.~Novikov, V.~A.~Novikov, A.~N.~Rozanov 
    and M.~I.~Vysotsky,
     hep-ph/9812346; \\
  A.~Dedes, A.~B.~Lahanas and K.~Tamvakis,
    \PRD{59}{99}{015019};
  G.~C.~Cho, K.~Hagiwara, C.~Kao and R.~Szalapski,
    hep-ph/9901351.
%
\bibitem{b-s-gamma_th}
  S.~Bertolini, F.~Borzumati, A.~Masiero and G.~Ridolfi,
   \NPB{353}{91}{591}; \\
  J.~Hewett,
   \PRL{70}{93}{1045}; \\
  V.~Barger, M.~Barger and R.~J.~N.~Phillips,
   \PRL{70}{93}{1368}; \\
  J.~L.~Lopez, D.~V.~Nanopoulos and G.~T.~Park,
    \PRD{48}{93}{974}; \\
  N.~Oshimo,
  \NPB{404}{93}{20}; \\
  F.~M.~Borzumati and F.~Vissani, 
   \ZPC{67}{95}{513}; \\ 
  J.~Wu, R.~Arnowitt and P.~Nath,
   \PRD{51}{95}{1371}; \\
  J.~L.~Lopez, D.~V.~Nanopoulos, X.~Wang and A.~Zichichi,
   \PRD{51}{95}{147}.
%
 \bibitem{peskin-takeuchi}
  M.~E.~Peskin and T.~Takeuchi,
   \PRL{65}{90}{964}; 
   \PRD{46}{92}{381}. 
%
 \bibitem{HHKM}
  K.~Hagiwara, D.~Haidt, C.~S.~Kim and S.~Matsumoto,
   \ZPC{64}{94}{559};
   \ZPC{68}{95}{352}(E).
%
 \bibitem{Mihara}
  S.~Mihara,
   ``Recent results from LEP-II'',
   talk given at KEK Theory Meeting on {\it Collider Physics},
   Jan 11-12, 1999 (Tsukuba, January 1999).
%
%
 \bibitem{Czarnecki}
   A.~Czarnecki, B.~Krause and W.~J.~Marciano,
   \PRL{76}{96}{3267}.
%
 \bibitem{Tevatron}
   D.~Amidei and R.~Brock, 
     ``Future electroweak physics at the Fermilab Tevatron, 
     Report of the TeV2000 Study Group : 
     Supersymmetric Physics'',
     FERMILAB-PUB-96-046 (1996).
%
 \bibitem{Davier}
  M.~Davier and A.~Hocker,
   \PLB{435}{98}{427}.
%
 \bibitem{Kuhn}
   J.~H.~Kuhn,
    Nucl.~Phys.~Proc.~Suppl.~{\bf 76}~(1999)~21.
%
 \bibitem{Alemany}
  S.~Eidelman and F.~Jegerlehner,
     \ZPC{67}{95}{585}; \\
%
  B.~Krause,
     \PLB{390}{97}{392}; \\
%
  R.~Alemany, M.~Davier and A.~Hocker,
    \EPJC{2}{98}{123}; \\
%
\end{thebibliography}
\end{document}